\documentclass[11pt]{article}

\usepackage{amsmath}
\usepackage{graphicx}
\usepackage{amsfonts}
\usepackage{amssymb}
\usepackage{epsfig}
\usepackage{color}
\usepackage{psfrag}

\newtheorem{rema}{Remark}[section]

\def\sect#1{\section{#1}\setcounter{equation}{0}}

\newcommand{\bc}{\begin{center}}
\newcommand{\ec}{\end{center}}
\def\ba#1{\begin{array}{#1}\displaystyle}
\newcommand{\ea}{\end{array}}

\newcommand{\beq}{\begin{equation}}
\newcommand{\eeq}{\end{equation}}
\newcommand{\beqa}{\begin{eqnarray}}
\newcommand{\eeqa}{\end{eqnarray}}

\newcommand{\n}{\nonumber\\}
\newcommand{\bi}{\begin{itemize}}
\newcommand{\ei}{\end{itemize}}

\def\lt#1{\left#1}
\def\rt#1{\right#1}

\def\h#1{\hat{#1}}

\def\frc#1#2{\frac{#1}{#2}}

\newcommand{\p}{\partial}

\newcommand{\Pexp}{{\cal P}\exp}
\newcommand{\vac}{{\rm vac}}
\newcommand{\bra}{\langle}
\newcommand{\ket}{\rangle}

\newcommand{\R}{{\mathbb{R}}}

\newcommand{\uH}{{\mathbb{H}}}

\newcommand{\Or}{{\cal O}}

\newcommand{\Tr}{{\rm Tr}}

\newcommand{\low}{{\mathbb{L}}}

\begin{document}

\bc {\bf \Large Nonequilibrium density matrix for thermal transport in quantum field theory}

{\em \'Ecole de physique des Houches:\\ Physique des syst\`emes quantiques fortement corr\'el\'es hors \'equilibre\\
Les Houches, Frances, 30 juillet - 24 ao\^ut 2012}
\vspace {0.4cm}

Benjamin Doyon\\
{\em 
Department of Mathematics\\
King's College London, U.K.}

\ec

In these notes I explain how to describe one-dimensional quantum systems that are simultaneously near to, but not exactly at, a critical point, and in a far-from-equilibrium steady state. This description uses a density matrix on scattering states (of the type of Hershfield's density matrix), or equivalently a Gibbs-like ensemble of scattering states. The steady state I am considering is one where there is a steady flow of energy along the chain, coming from the steady draining / filling of two far-away reservoirs put at different temperatures. The context I am using is that of massive relativistic quantum field theory, which is the framework for describing the region near quantum critical points in any universality class with translation invariance and with dynamical exponent $z$ equal to 1. I show, in this completely general setup, that a particular steady-state density matrix occurs naturally from the physically motivated Keldysh formulation of nonequilibrium steady states. In this formulation the steady state occurs as a result of a large-time evolution from an initial state where two halves of the system are separately thermalized. I also show how this suggests a particular dependence (a ``factorization'') of the average current on the left and right temperatures. The idea of this density matrix was proposed already in a recent publication with my collaborator Denis Bernard, where we studied it in the context of conformal field theory.

\tableofcontents

\sect{Introduction}

These notes will explain how to describe quantum systems that are simultaneously near to (but not exactly at) a quantum critical point, and in certain steady states away from equilibrium. This situation is particularly intriguing because it mixes in an essential fashion both the rich collective quantum effects and nonequilibrium physics. Also, steady states, being time independent, are as near as possible to the usual states of equilibrium physics while still being out of equilibrium, and we can provide general descriptions in parallel to parts of equilibrium statistical physics.

We have in mind quantum systems composed of a macroscopic number of microscopic constituents with local (few-neighbors) interactions on some lattice. In certain situations, for instance when a parameter (a magnetic field, an interaction strength, etc.) is adjusted appropriately, the system becomes critical. Near to such a quantum critical points, quantum effects dominate the low-energy and large-distance physics. Indeed, on the one hand, the ground state encodes quantum correlations between local observables that exist much beyond few lattice sites. This is nontrivial, because with local interactions, one expects that constituents separated by many sites do not ``feel" each other very much, which is indeed what happens away from critical points. On the other hand, at least intuitively, the low-lying excitations can be understood as large-scale collective behaviors of the underlying constituents that emerge thanks to quantum fluctuations; as if large groups started to act together forming new quantum objects. Critical points arguably give some of the most interesting physics: they are where many nontrivial physical effects arise (one of the most well-known being the Kondo effect); and their large-distance / low-energy physics is actually universal, because most of the intricacies of the local interactions have been lost in the averaging out over many sites that arises in large-distance correlations or collective objects.

The physical theory that describes what is observed ``near'' critical points is quantum field theory (QFT) (more precisely, when not exactly at the critical point, massive QFT). By ``near'' we mean in the limit where observation lengths, lengths related to the system's geometry and correlation lengths are all large in lattice spacings, these three kind of lengths being taken to infinity in fixed proportions\footnote{The concept of observation length is kept slightly nebulous here, but as examples this could be the distance between two points in a correlation function; or, translating into the energy-momentum language by the usual Fourier-transform inversion, the temperature.}. This is the scaling limit. QFT is the physical theory that will provide predictions for what happens in the scaling limit (for instance, how correlations decay to zero, giving both the exponents and universal amplitudes). Of course, in reality one never reaches exactly the scaling limit. But when criticality is approached in macroscopic systems, then every large-distance and low-energy observations are effectively predicted by QFT.

Note that if the correlation lengths are much larger than everything else, then we are exactly ``at'' the critical point. In these cases, QFT possesses scale invariance, and often with this, invariance under the full global conformal group. In 1+1 dimensions (the scaling limit of quantum chains), this very often comes with local conformal invariance, and we have a conformal field theory (CFT). We will not describe these situations here, and refer the reader to \cite{BD12} for the CFT counterpart of the nonequilibrium steady states we will study here.

The equilibrium physics of critical and near-critical points has been very well studied over the years, and there is by now a wide range of techniques to obtain exact and approximate results. Not as much can be said of the physics away from equilibrium. One reason is, of course, the large variety of out-of-equilibrium situations (instead of saying what the situation is, we say what it is not...). But also, even in the relatively simple cases where a steady state exists, there lacks a framework as powerful as and as widely applicable as that of Gibbs ensembles. Hence, one often must revert to other formulations, usually rather adapted to the type of physical model at hand. Here, we will start, in the massive QFT context, with an ``explicit'', real-time construction of the steady state, where we implement theoretically the idealized experimental set-up that should produce the steady state\footnote{Similar real-time constructions, where unitary evolution is performed from a state that is not a Hamiltonian eigenstate, are usually referred to, in general, as ``quenches.''}. This will lead us to rather general statements, paralleling Gibbs ensembles.

\sect{Basic notions} \label{sectss}

Consider a quantum chain of length $L$ with local (nearest-neighbor, say) interactions. We want to describe a state where energy flows along the chain due to a difference of temperatures at its extremities. There are various ways to do this: one could use Lindblad's theory of open quantum systems where the thermal baths have been traced out and the time evolution is non-unitary \cite{Lind}, or, in a somewhat related fashion, one could explicitly connect Caldeira-Leggett baths \cite{CaldeiraL81} with different temperatures on the extremities of the chain (as is done, for instance, in \cite{SD07,SD11}). One also could consider an infinite chain, instead of a finite one, and try to describe the flow using asymptotic states in a Gibbs-like ensemble, through what is usually referred to in the condensed matter community as Hershfield's density matrix (Hershfield considered this for general charge transport \cite{Hershfield93}; Ruelle obtained rigorous general results in the context of $C^*$-algebras with similar underlying ideas for heat transport \cite{R00}, see also \cite{AJPP06}); this of course requires a clear principle to fix this density matrix. Finally, also with boundaries sent to infinity, one could explicitly construct the baths with different temperatures by using parts of the system itself as baths: the infinite chain is initially cut into two semi-infinite parts, each thermalized at different temperatures, which are then connected and evolved unitarily for a long time. This latter method is in a sense the most ``physical'', as the steady state is explicitly constructed in real time (it is sometimes referred to as the ``partitioning approach'' of Caroli et al. \cite{C71}, and sometimes as the Keldysh formulation, used for instance in the context of mesoscopic systems \cite{JWM94}, and it is related to quantum quenches). Note that in this method, thanks to the unitary evolution, no part of the system is kept at a fixed temperature during time evolution; the information of the temperature is only in the initial mixed state. When combined with ideas of perturbation theory and path integrals, the Keldysh formulation leads to the powerful Keldysh perturbation theory (in fact, Keldysh perturbation theory can be used as well with the method of the Cladeira-Leggett baths).

In the present notes, we will start with the Keldysh formulation in the context of general, nonperturbative massive QFT, and we will show, using non-rigorous but well motivated QFT arguments, that it leads to a Hershfield formulation, giving the precise Gibbs-like ensemble of asymptotic states describing the steady state where there is an energy flow. This was already proposed in \cite{BD12}, where it was studied in the context of CFT (there, the heat current and its full counting statistics were found). See also \cite{AP03,DA06,H07,D09,DKHH11,MCP11,BD12b,DSH11,BD12} for similar constructions concerning charge or heat transport in a variety of other situations and at various levels of rigor.

To be more specific, we consider the Hamiltonians $H^l$ and $H^r$ of the two semi-infinite, separate halves (left and right) of a quantum chain, and the initial density matrix
\beq\label{rhoo}
	\rho_0 = e^{-\beta_lH^l - \beta_rH^r}
\eeq
where the halves are at different temperatures $T_{l,r}=\beta_{l,r}^{-1}$. We then evolve this density matrix with the full Hamiltonian $H=H^l+H^r + \Delta H$ where $\Delta H$ represents the energy term for the single link connecting both halves:
\beq
	\rho(t) = e^{-iHt}\rho_0 e^{iHt}.
\eeq
The steady state average of observables is obtained upon evaluating $\lim_{t\to\infty} \Tr\lt(\rho(t)\Or\rt)/\Tr\lt(\rho(t)\rt)$. More precisely, if $v$ is a typical velocity of quantum propagation in the chain (which we will take below to be 1), then the limit is that $vt$ must be much greater than the observation length (which could be taken as the length of the support of the observable $\Or$), and much smaller than the length of the system itself (which we have implicitly taken as infinity). This is so that both halves of the system are big enough to play the role of thermal baths, absorbing and emitting energy as necessary, so that a nonequilibrium steady state may arise. See \cite{DA06,D09,BD12b} for discussions about such steady-state limits.

\begin{rema}
One may worry about the physical meaning, in the Hershfield and Keldysh formulations, of sending the extremities of the chain to infinity. Indeed, according to Fourier's law, the heat current (which is, in many situations, the same as the energy current) is proportional to the gradient of the ``local temperature'': $j = -c\nabla T$ with heat conductivity $c$. But if the extremities are infinitely far from each other, and no part of the system is actually kept at a fixed temperature during the evolution, surely the local temperature gradient, after infinite time, will be zero, hence there should be no heat current, unless the heat conductivity $c$ is infinite. Yet one of course find that the heat current is nonzero (although we will not evaluate any current in the present notes). We must realize that in low dimensions, Fourier's law is known to be broken: heat conduction is anomalous \cite{BLR00,LLP03,D08}, and gets non-local contributions. In particular, it was shown in \cite{KIM12} that a nonzero Drude weight plays an important role. An underlying principle in the present context may be that of quantum coherence. Returning to Fourier's law for the sake of the argument, it should be understood in a coarse-grained fashion: one must take the limit of $T(x+\Delta x)/\Delta x$ where $\Delta x$ is much smaller than the distance at which $T$ changes significantly, but much greater than the distance at which things start to mix and scatter (so that, in particular, the concept of a local temperature is clear). But quantum coherence precludes scattering and mixing; so $\Delta x$ must be much greater than the quantum coherence length. Near to quantum criticality, the correlation length is very large (in lattice spacings), hence one may indeed expect the quantum transport behavior to be very different from that predicted by Fourier's law.
\end{rema}

\begin{rema}\label{remexlimits}
We will consider the case where there is a quantum critical point, and take the scaling limit, described by QFT. See \cite{BD12b,BD12c} for a discussion of the interplay between the scaling limit and the steady-state limit. In particular, the scaling limit describes the low-energy behaviors, and one may worry that some low-energy states of $H^l+H^r$ are high energy states of $H$ and are not captured by a QFT description of the steady-state limit (i.e.~lie far outside the low-energy, universal range). The basic explanation \cite{BD12b,BD12c} is that if the renormalization of the observable with respect to the $H^l+H^r$ ground state is also finite with respect to the $H$ ground state, then the correct steady-state average of this observable is captured by QFT.
\end{rema}

\begin{rema}
Instead of connecting a single link, we could also think of connecting a whole finite-size (in lattice spacing) subpart of the chain (i.e.~$\Delta H$ corresponds to a finite number of sites). Near to critical points, this will lead to the same low-energy results, because the correlation length is infinite in lattice spacings. In fact, we also expect that with $\Delta H$ corresponding to a number of site being a finite proportion of the correlation length, we still find the same steady state.
\end{rema}



\sect{Energy-flow steady states in massive 1+1-dimensional quantum field theory}\label{sectmain}

We restrict our attention to translation-invariant quantum chains in the thermodynamic limit and assume that there is a quantum critical point with dynamical exponent $z=1$. Averages near to, or at, quantum critical points have asymptotic behaviors described by QFT. With $z=1$, the near-critical (not exactly critical) behavior of the quantum chain is described by massive relativistic QFT, which is the context that we consider. See for instance the standard reference \cite{Sach} for a description of quantum critical points. In essence, if $m$ is the gap between the lowest energy state of the quantum chain and the next state, and $J$ is some microscopic energy scale (like the intersite coupling), then the scaling limit leading to massive QFT is that where $T_l,T_r <,\simeq m \ll J$.

\subsection{Asymptotic states and field configurations}

One way of thinking about QFT is as a theory for quantized field. Although this is exactly what the name relates to, it is not entirely accurate because in particular of the need for renormalization; yet it does provide a good intuitive picture, and gives rise to precise results when used correctly. In this picture, QFT is a quantum theory whose Hilbert space ${\cal C}$ has for basis the set $\{|\phi\ket\}$ parametrized by all field configurations $\phi$ on real space, with overlap $\bra\phi|\phi'\ket = \delta[\phi,\phi']$ the delta-functional supported on the field configurations being equal\footnote{It is not clear that this is actually a Hilbert space: restriction to the set of field configurations may be necessary, beside other subtleties; but for our intuitive description, this is sufficient.}. In the 1+1-dimensional case, these are field configurations $\phi:\R\to {\cal S}$ on the line, where the field takes values in some space ${\cal S}$. There is the associated $x$-dependent operator $\h\phi(x)$ (which, following tradition, we refer to as a field) defined by $\h\phi(x)|\phi\ket = \phi(x)|\phi\ket$, and there is an energy density $\h h(x)$, formed out of $\h\phi(x)$ and its derivatives, using which the Hamiltonian is
\beq
	H = \int_{-\infty}^\infty dx\,\h h(x).
\eeq
We can think intuitively of a field configuration as being the conglomeration of the individual basis states that the microscopic components of the underlying quantum model are in; but we must realize that on any finite length on $\R$, there are infinitely many such components, and that in the effective description of QFT, the space ${\cal S}$ is not necessarily the set of such microscopic-component basis states.

A somewhat surprising consequence of the principles of QFT (locality and Poincar\'e invariance) is that the Hamiltonian eigenfunctions and eigenvalues are completely characterized by few parameters: the particle spectrum. Indeed, given a set of quantum numbers ${\cal N}$ and associated particle masses ${\cal M} = \{m_\alpha:\alpha\in {\cal N}\}$, a set of orthonormal basis states in the Hilbert space is given by
\beq\label{basis}
	\{|\theta_1,\ldots,\theta_k\ket_{\alpha_1,\ldots,\alpha_k}:
	\theta_1>\ldots>\theta_k,\;\alpha_j\in {\cal N},\;k=0,1,2,\ldots\}
\eeq
on which the Hamiltonian has eigenvalues
\beq
	\sum_{j=1}^k m_{\alpha_j}\cosh\theta_j.
\eeq
Here we are using the rapidity $\theta$, in terms of which the relativistic (on-shell) energy and momentum are given by $E = m\cosh\theta$ and $p = m\sinh\theta$. When $k=0$, we denote the state by $|\vac\ket$. The sets ${\cal N}$ and ${\cal M}$ are characteristics of the QFT model, although they do not completely define it.

\begin{rema}
Hence, in massive relativistic QFT, the problem of diagonalizing the Hamiltonian is almost solved: one only needs to determine few parameters! Determining the masses can be hard of course, and even when this is done, the QFT model is not yet solved, neither even fully defined. One more piece of information is necessary: that having to do with locality. One needs to say what the stress-energy tensor is, as a (tensorial) operator. This not only provides a definition of space via the momentum operator, but also of what operators are local: one can write the quantum locality condition of vanishing commutators with the energy density at space-like distances. This fully defines the QFT model, and to solve it (according to a certain definition), one needs to find all local operators and calculate their correlation functions. This in principle provides the measurable physical information encoded in the QFT model.
\end{rema}

The basis (\ref{basis}) is a basis of asymptotic states. Naturally, asymptotic states are particular linear combinations of field configurations, formally
\beq
	|\theta_1,\ldots,\theta_k\ket_{\alpha_1,\ldots,\alpha_k}
	= \sum_\phi c_{\alpha_1,\ldots,\alpha_k}
	(\,\theta_1,\ldots,\theta_k\,|\,\phi)\, |\phi\ket.
\eeq
Intuitively, they are linear combinations such that if we were to evolve the state, say, back in time for a very long time, we would find large wave packets widely separated and propagating at rapidities $\theta_1,\ldots,\theta_k$. The wave packets would be associated to the various particle types $\alpha_1,\ldots,\alpha_k$, so they would be wave packets of the various ``fundamental fields'' related to the various particle types. These various fundamental fields can be components of $\h\phi(x)$ (remember that $\phi$ takes values in ${\cal S}$, which may be $\R^n$ for a $n$-component field for instance), or some composite fields (like $\h\phi(x)^2$, appropriately renormalized), depending on the model.

Of course, the paragraph above cannot be entirely true, because asymptotic states diagonalize the Hamiltonian; hence they evolve trivially (only getting a phase). A more correct statement is that one must construct linear combinations with the properties of the above paragraph, and then take the limit of infinite time, simultaneously infinitely-large and infinitely-separated wave packets. In order to represent scattering states, this limit must be taken in such a way that the ``free'' relativistic evolution of wave packets would produce space-time trajectories (for the centers of the wave packets) that are found, at time 0, in a finite region of space. Where exactly they are found (so-called impact factors) is up to us, and just affect our choice of basis elements.

\begin{rema}
What is described here are in-states. There is a similar construction where the condition is obtained by evolving forward in time; the resulting asymptotic states are out-states. These then provide two bases for the Hilbert space that diagonalize the energy operator; they have the same description and lead to the same energies, but are different bases. The overlaps between in-states and out-states forms the scattering matrix of the model. It is expected that giving the scattering matrix is equivalent to specifying the stress-energy tensor: it provides the locality information of the QFT model. This is rather explicit in the context of integrable QFT, with the so-called form factor program, but not obvious out of integrability.
\end{rema}

\begin{rema}\label{remacompletion}
Not all possible linear combinations $\sum_\phi c_\phi |\phi\ket$ of field configurations occur in the basis (\ref{basis}) (or in finite linear combinations of basis elements); in particular the states $|\phi\ket$ themselves do not occur. But we expect that by taking the completion, involving all (possibly infinite, and continuously so) linear combinations that converge, one recovers an appropriate set of field configurations $|\phi\ket$ and their (again possibly infinite, and continuously so) linear combinations $\sum_\phi c_\phi |\phi\ket$ with well-behaved coefficients $c_\phi$. In fact, we expect to obtain all square-integrable field configurations $\phi$ -- this is natural, as $\phi$ is supposed to represent a wave function in the ``second-quantization'' viewpoint on QFT.  This is similar to the situation in ordinary quantum mechanics, where the wave functions for energy eigenstates may have certain continuity and differentiability properties, and appropriate infinite linear combinations need to be taken in order to obtain all possible square-integrable functions, including for instance the discontinuous ones. In the present case, however, this  is much more complicated, because the basis (\ref{basis}) has continuous components (it is not a countable basis); hence the full Hilbert space structure (including completeness) is much more subtle. We will avoid these tricky details here, and only discuss basic properties in due course.
\end{rema}

It is extremely important to realize that asymptotic states are just particular linear combinations of field configurations. The term ``asymptotic'' refers here to the way these linear combinations are chosen: by an asymptotic condition on a hypothetical time evolution. It does not refer to the states ``living only in the far past'' (or the far future). It turns out that this construction produces linear combinations of field configurations that diagonalize the Hamiltonian.

We can verify the latter sentence and make the intuitive construction above much more precise as follows. Recall first that the vacuum in quantum field theory is itself a nontrivial linear combination of field configurations:
\beq\label{vac1}
	|\vac\ket = \sum_{\phi} c(\vac|\phi)\,|\phi\ket.
\eeq
One standard way of defining this linear combination is via the path integral (appropriately renormalized). Assume that the Hamiltonian is associated to the action functional $S[\varphi]$, for $\varphi:\R^2\to {\cal S}$, where $\varphi(x,t)$ is the trajectory in time $t$ of the field configuration $\phi$. From this it is of course possible to write down the action $S_R[\varphi]$ on any open region $R$ of $\R^2$. One defines the coefficients $c(\vac|\phi)$ by
\beq\label{vac2}
	c(\vac|\phi) = \int_{\varphi(x,0^-) = \phi(x)\;:\;x\in\R\atop \varphi(x,t)\to0\;:\;x\in\R,\,t\to-(1-i0^+)\infty}
	{\cal D}_\low \varphi\,e^{iS_\low[\varphi]}
\eeq
where $\low = \{(x,t)\in\R^2:t<0\}$ and ${\cal D}_\low\varphi$ is an integration measure over all configurations of $\varphi:\low\to{\cal S}$. That is, one performs Feynman's path integral on field trajectories over half of space-time, the half with negative times, with the boundary condition in the limit of zero time $t\to0^-$ given by the field configuration $\phi$, and with infinite-past asymptotic condition (under Feynman's imaginary-time prescription) equal to 0 (here we assume that $\phi=0$ is a minimum of the potential representing the interaction).

The wave packets are constructed on top of this vacuum. Let us denote by $\h\phi_\alpha(x)$ a fundamental field associated to the particle $\alpha$. This means that $\h\phi_\alpha(x)$ has the quantum numbers of the particle $\alpha$, and that the Fourier transform of the two-point correlation function
\beq
	\bra\vac|{\cal T}\lt(\h\phi_\alpha(x,t)\h\phi_\alpha(0,0)\rt)|\vac\ket
\eeq
(where ${\cal T}$ is the time-ordering operation, bringing the earliest operator to the right, and $\h\phi_\alpha(x,t) = e^{iHt}\h\phi_\alpha(x)e^{-iHt}$) has a pole, as function of the square two-momentum $E^2-p^2$, at $m_\alpha^2$ and no other singularities at lower values. We assume for simplicity of the discussion that $\h\phi_\alpha$ is a real bosonic field, transforming under the Poincar\'e group in the spin-0 representation (generalizing to other representations and statistics and to complex fields is straightforward, if somewhat tedious). The statement above implies that the correlation functions of $\h\phi_\alpha(x,t)$ with any other local fields satisfy the Klein-Gordon equation asymptotically in time (other equations occur for other representations):
\beq\label{KG}
	\lt(m^2-\frc{\p^2}{\p x^2} - \frc{\p^2}{\p t^2}\rt) \bra {\cal T}\lt(
	\h\phi_\alpha(x,t)\cdots\rt)\ket= O\lt(e^{-m'\sqrt{t^2-x^2}}\rt)\quad
	\mbox{as} \quad t^2-x^2\to\infty.
\eeq
Here $m'$ is the lowest mass, greater than $m$, created by $\phi_\alpha$, and $\cdots$ represents other fields at fixed positions.

Define the following operators:
\beq\label{A}
	A_\alpha(\theta) =
	i\lim_{t\to-\infty\atop L\to\infty}\int_{-\infty}^\infty dx
	\lt(f_\theta(x,t) \frc{\p}{\p t}\h\phi_\alpha(x,t) -
	\frc{\p}{\p t}f_\theta(x,t)
	\h\phi_\alpha(x,t)\rt)
\eeq
with
\beq\label{wavepacket}
    f_\theta(x,t) = \exp\lt[im\cosh(\theta)\,t - im\sinh(\theta)\,x -
    \frc{\lt(x-\coth(\theta)t\rt)^2}{L^2}\rt]~,
\eeq
as well as their Hermitian conjugate $A_\alpha^\dag(\theta)$. Here the limit on $t$ and $L$ is such that $t\gg L$, so that the wave packets are well separated (without going into more details). Using (\ref{KG}) and taking into account the limits taken in (\ref{A}), one can show that these are eigenoperators of the Hamiltonian, in particular:
\beq
	[H,A^\dag_\alpha(\theta)] = m_\alpha\cosh\theta \,A^\dag_\alpha(\theta).
\eeq
The vacuum having zero energy, this means that we can construct eigenstates of the Hamiltonian by multiple actions of $A^\dag_\alpha(\theta)$:
\beq\label{asA}
	|\theta_1,\ldots,\theta_k\ket_{\alpha_1,\ldots,\alpha_k}
	= A^\dag_{\alpha_1}(\theta_1)\cdots A^\dag_{\alpha_k}(\theta_k)
	|\vac\ket.
\eeq
These are the asymptotic (in-)states.

One may wonder why we didn't use the operators $A_\alpha(\theta)$ instead in order to create different Hamiltonian eigenstates out of the vacuum. This is because
\beq\label{vacA}
	A_\alpha(\theta)|\vac\ket = 0
\eeq
and
\beq\label{ccr}
	[A_{\alpha}(\theta), A_{\alpha'}^\dag(\theta')] = 4\pi\delta_{\alpha,\alpha'}\delta(\theta-
	\theta').
\eeq
The former is a (slightly subtle) consequence of (\ref{vac1}), (\ref{vac2}) and (\ref{A}), (\ref{wavepacket}). The latter is more involved, and a consequence of locality and of the pole structure of the Fourier transforms of the two-point functions $\bra\vac|{\cal T}\lt(\h\phi_\alpha(x,t)\h\phi_{\alpha'}(0,0)\rt)|\vac\ket$. Relations (\ref{vacA}) and \ref{ccr}) mean that the vacuum (\ref{vac1}) can alternatively be defined as the vacuum of the Fock space of the canonical algebra generated by $A_\alpha(\theta)$ and $A^\dag_\alpha(\theta)$; this is a definition which does not rely on the existence of an action functional.

\begin{rema}
Note that the canonical commutation relations (\ref{ccr}) do not imply that the model is free. They only relate to the fact that the asymptotic states have a particularly simple description, which can be obtained as the Fock space of a canonical algebra. But for interacting models, the local fields are not simple Fourier transforms of the operators $A_\alpha(\theta)$, $A_\alpha^\dag(\theta)$, and taking the limit $t\to\infty$ instead of $t\to-\infty$ in (\ref{A}) gives rise to operators creating the out-states, which are in general different form the in-states.
\end{rema}

Out of the states (\ref{asA}), we may now construct the Hilbert space of the model by completion. Without going into any detail of this rather subtle mathematical process (see Remark \ref{remacompletion}), we will denote by ${\cal H}$ the resulting Hilbert space; this is expected to be a proper subspace of the space of field configurations ${\cal C}$.

\subsection{Asymptotic states in the presence of a cut-impurity}

Impurities and boundaries on the quantum chain give rise to QFT models with inhomogeneities running, in space-time, along the time direction (hence there is a natural, preferred time direction). These can be treated in similar ways as above. The case of interest below is that where there is a cut at one point along the chain (that is, we have two separate half-infinite chains). In this case, the field configurations $\phi$ are on $\R^0:=\R\setminus \{0\}$ (the basis $|\phi\ket$ forms the space ${\cal C}^0$), and the Hamiltonian $H^0$ is (slightly) different from $H$, in its expression in terms of fields $\h\phi$, having a discontinuity at $x=0$:
\beq
	H^0 = \int_{\R^0} dx\,\h h(x).
\eeq
Obviously, we can write $H^0 = H^l + H^r$ as a sum of a term on $(-\infty,0)$ and a term on $(0,\infty)$, respectively. These terms commute with each other, $[H^l,H^r]=0$, hence can be diagonalized simultaneously.

The vacuum can be described as
\beq\label{vaco1}
	|\vac\ket^0 = \sum_{\phi} c^0(\vac|\phi)\,|\phi\ket
\eeq
where
\beq\label{vaco2}
	c^0(\vac|\phi) = \int_{\varphi(x,0^-) = \phi(x)\;:\;x\in\R^0\atop \varphi(x,t)\to0\;:\;x\in\R^0,\,t\to-(1-i0^+)\infty}
	{\cal D}_{\low^0} \varphi\,e^{iS_{\low^0}[\varphi]}
\eeq
with $\low^0 = \{(x,t)\in\R^2:x\neq 0, t<0\}$. Here we choose the free boundary conditions as $x\to0^-$ and $x\to0^+$, hence we do not constrain the limit value of the field. This is clearly a factorized vacuum, $|\vac\ket^0= |\vac\ket^l\otimes |\vac\ket^r$.

Then, asymptotic states are constructed in the same way as for the model on the whole line. We define $A_\alpha^0$ by
\beq\label{A0}
	A_\alpha^0(\theta) =
	i\lim_{t\to-\infty\atop L\to\infty}\int_{-\infty}^\infty dx
	\lt(f_\theta(x,t) \frc{\p}{\p t}\h\phi_\alpha^0(x,t) -
	\frc{\p}{\p t}f_\theta(x,t)
	\h\phi_\alpha^0(x,t)\rt)
\eeq
with $\h\phi_\alpha^0(x,t) = e^{iH^0t}\h\phi_\alpha(x)e^{-iH^0t}$.
Then we simply act on the vacuum with the operators $(A_\alpha^0)^\dag(\theta)$:
\beq\label{asA0}
	|\theta_1,\ldots,\theta_k\ket_{\alpha_1,\ldots,\alpha_k}^0
	= (A^0_{\alpha_1})^\dag(\theta_1)\cdots (A^0_{\alpha_k})^\dag(\theta_k)
	|\vac\ket^0.
\eeq
The $l$-$r$ factorization also occurs for asymptotic states. Note that for positive rapidities, the wave packets (\ref{wavepacket}) are supported on $(-\infty,0)$, and for negative rapidities, on $(0,\infty)$ (apart from exponentially decaying tails which vanish in the limit $t\to-\infty$). Since asymptotic state operators commute for rapidities of different signs, we can then write
\beq
	|\theta_1,\ldots,\theta_k\ket_{\alpha_1,\ldots,\alpha_k}^0
	= \prod_{i:\theta_i>0}(A_{\alpha_i}^0)^\dag(\theta_i) |\vac\ket^l
	\otimes
	\prod_{j:\theta_j<0}(A_{\alpha_j}^0)^\dag(\theta_j) |\vac\ket^r.
\eeq
The factors are the separate eigenstates of $H^l$ and $H^r$.

Again, it is important to recall that the asymptotic states (\ref{asA0}) are just particular linear combinations of field configurations on $\R^0$. In particular, the field configurations involved, seen as functions on $\R$, are generically discontinuous at $x=0$ (that is, the left- and right- limits generically don't agree). Despite the similar formulation, the asymptotic states (\ref{asA0}) are very different from the asymptotic states (\ref{asA}), when seen as linear combinations of field configurations (in particular, there is reflection on the point $x=0$ in the case of the theory on $\R^0$).

\begin{rema}
A model on $\R^0$ is equivalent to two independent copies of the model on a half-line. In each copy, the set of asymptotic states is twice smaller than that for the model on the line, and one can describe them by choosing a sign of the rapidities. This is clear from the asymptotic state construction above: for instance, only states where particles have positive momenta have wave packets supported inside $(-\infty,0)$, as noted. Putting two copies together, we get a set of asymptotic states isomorphic to that of the model on the line. There are different ways of describing the resulting tensor product states. But if one copy is interpreted as being on $(-\infty,0)$ and the other on $(0,\infty)$ (so that we have a model on $\R^0$, as we did above), then we have both signs of rapidities involved, and the asymptotic state construction that we have gives the same description (\ref{basis}) both for the set of asymptotic states on $\R$, and that on $\R^0$. Here we neglect states with particles of zero rapidities, because they contribute only infinitesimally to averages of local operators.
\end{rema}

\begin{rema}
Impurities and boundaries give rise to more intricate scattering phenomena, which include in general reflection and transmission phenomena. This is not seen in the construction of asymptotic in-states, or of asymptotic out-states; it is rather seen in the structure of the scattering matrix, formed by the overlaps between in- and out-states. The scattering matrix will have additional components corresponding to reflections and transmissions. In the case of the cut-impurity, of course, only reflection phenomena will occur at $x=0$, without transmission. In general, one could also have additional bound states to the impurity; but this is not expected to occur with a cut impurity.
\end{rema}

Out of the states (\ref{asA0}), we may now construct the Hilbert space ${\cal H}^0 = {\cal H}^l\otimes {\cal H}^r$ of the model by completion. We omit any discussion of the subtleties of how ${\cal H}^0$ compares to ${\cal H}$ (e.g.~how it is embedded into ${\cal H}$), but note that the two Hilbert spaces are closely related, since only one point on the line is missing in the construction of ${\cal H}^0$. From the viewpoint of the asymptotic state basis, the only difference between ${\cal H}$ and ${\cal H}^0$ is of course the states with zero-rapidity particles, which, as mentioned, are expected to contribute only infinitesimally to averages of local (or finite-extent) observables.



\begin{rema}\label{remaHilbertspace}
It is not essential to single out the point $x=0$ as we have done above, since in QFT, the value of the field at one point in space does not affect physical results (unless infinite potentials exist at that point). From this viewpoint, we may expect ${\cal H}^0\cong {\cal H}$, and we could simply add the point at $x=0$ from a configuration on $\R^0$ in some more or less arbitrary fashion (e.g.: the average of the right and left limits). However, singling out $x=0$ clarifies the discussion and avoids these technical details. We may also think of the point $x=0$ to be representing the scaling limit of some finite, contiguous-sites subsystem of the underlying quantum chain that separates the half-infinite left and right parts. Our results apply to this general situation. In fact, we may replace the point $x=0$ by any finite interval $[a,b]$, $a,b\in\R$. On the left / right of the interval, we may use the description of ${\cal H}^l\otimes {\cal H}^r$ as above, and we would have an additional space ${\cal L}$ to describe the configurations on the interval itself. Our main results would still hold in this case.
\end{rema}

\subsection{Steady-state density matrix for energy flows}

\subsubsection{Statement of the result}

According to the steady-state construction in Section \ref{sectss}, and taking the scaling limit, we need to evaluate the average of observables $H$-evolved to time $t$, in the mixed state with density matrix $\rho_0$ given by (\ref{rhoo}). The left half of the system, with energy $H^l$, is at temperature $\beta_l^{-1}$, and the right half, with energy $H^r$, is at temperature $\beta_r^{-1}$. The steady state is obtained in the limit where $t\to\infty$. For our observable, we will take any operator $\Or$ that is {\em finitely-supported}. Formally, this is an operator that commutes with the energy density $\h h(x)$ whenever $|x|$ is large enough; informally, it is composed of local fields (in products, integrals) at points that all lie on some finite closed interval (the support). This guarantees that far enough in space, the observable has no effect.

We will show that
\beq\label{toshow}
	\lim_{t\to\infty} \frc{\Tr_{{\cal H}^0}\lt[\rho_0  \,e^{iHt} \Or e^{-iHt}\,\rt]}{\Tr_{{\cal H}^0}[\rho_0]} =
	\frc{\Tr_{{\cal H}}\lt[\rho_{\rm stat} \Or \rt]}{\Tr_{{\cal H}}[\rho_{\rm stat}]}
\eeq
for every finitely-supported operator $\Or$. Here, $\rho_0$ is diagonal on the asymptotic-state basis (\ref{asA0}) of ${\cal H}^0$ with eigenvalues
\beq\label{elr}
	e^{-\beta_l E^l-\beta_r E^r}
\eeq
where $E^l$ and $E^r$ are the eigenvalues of $H^l$ and $H^r$, respectively; and $\rho_{\rm stat}$ is diagonal on the asymptotic-state basis (\ref{asA}) of ${\cal H}$ with eigenvalues
\beq\label{epm}
	e^{-\beta_l E^+ - \beta_r E^-};
\eeq
where
\beq\label{lrpm}
	E^+ =E^l = \sum_{i:\theta_i>0} m_{\alpha_i} \cosh\theta_i,\quad
	E^- =E^r = \sum_{j:\theta_j<0} m_{\alpha_j} \cosh\theta_j.
\eeq
Note that $E^+$ and $E^-$ have the same form as $E^l$ and $E^r$, respectively, when expressed in terms of the rapidities of asymptotic states. However, we emphasize that the asymptotic states associated to the values $E^\pm$ are very different from those associated to the values $E^{l,r}$.

Using the density operators $n_\alpha(\theta) = A^\dag_\alpha(\theta) A_\alpha(\theta)$, which measure the density of asymptotic particles of type $\alpha$ at rapidity $\theta$, we can write explicitly $\rho_{\rm stat}$ as
\beq
	\rho_{\rm stat} = \exp\lt[-\beta_l\int_{0}^{\infty} d\theta\,
	\sum_\alpha n_\alpha(\theta)
	-\beta_r\int_{-\infty}^{0} d\theta\,
	\sum_\alpha n_\alpha(\theta)\rt].
\eeq
Hence we have a description of the non-equilibrium steady state via a density matrix; a ``quasi-equilibrium'' description.

\begin{rema}
Note that $E^++E^-$ is the eigenvalue of $H$ on the state (\ref{asA}), but that $\beta_l E^+ + \beta_r E^-$ is not, in general, the eigenvalue of a local charge (the integration of a local density). This is in contrast to the situation on ${\cal H}^0$: not only $E^l+E^r$ is the eigenvalue of $H^0$, but also $\beta_l E^+ + \beta_r E^-$ is the eigenvalue of $\beta_l H^l + \beta_r H^r$, a local charge. Nonlocality of the charge involved in the nonequilibriun steady-state density matrix is a quite general phenomenon, observed also in the cases of charge currents in impurity models \cite{DA06,D07}.
\end{rema}

\begin{rema}\label{rematrace}
Although we discuss density matrices and traces over them, there are a great many subtleties involved into a proper definition of such objects. One way to define them is to regularize to a finite-length quantum chain in the same universality class, and then take the limit after evaluating the trace. This is correct but in general rather messy, and the difficulty in taking the scaling limit makes it hard to extract the universal structures associated to QFT from the particularities of the quantum chain chosen. Another way is to consider the operator algebra of QFT independently of the states ($C^*$-algebra), and to see the trace as a particular linear functional on this algebra. This is the algebraic-QFT way, is mush more elegant, and can be directly applied to the infinite-length limit, which is useful given that this limit must be taken first in the construction of the steady state. Unfortunately, it is hard to do anything else than the free theory (and aspects of conformal field theory) in this completely rigorous picture. Taking ideas from both perspectives may be the most fruitful way.
\end{rema}

The derivation of (\ref{toshow}) is based on the statement that the operator $e^{-iHt} e^{iH^0t}$, in the large-$t$ limit, is a map ${\cal H}^0\to {\cal H}$ which takes the state (\ref{asA0}), times any state on ${\cal L}$, to the state (\ref{asA}):
\beq\label{mapstates}
	\lim_{t\to\infty}e^{-iHt} e^{iH^0t}
	|\theta_1,\ldots,\theta_k\ket_{\alpha_1,\ldots,\alpha_k}^0
	= |\theta_1,\ldots,\theta_k\ket_{\alpha_1,\ldots,\alpha_k}.
\eeq
This statement is true in the context of evaluating matrix elements of finitely-supported operators (that is, in the ``weak'' sense); in particular, there is no problem with a possible loss of invertibility after the limit process (see Remark \ref{remaHilbertspace}). This statement shows (\ref{toshow}), because, $H^0$ commuting with $\rho_0$, we have
\beq\label{thh}
	\frc{\Tr_{{\cal H}^0}\lt[\rho_0  \,e^{iHt} \Or e^{-iHt}\rt]}{\Tr_{{\cal H}^0}[\rho_0]}
	=
	\frc{\Tr_{{\cal H}^0}\lt[\rho_0 \,e^{iHt}e^{-iH^0t}
	 \Or e^{-iHt}e^{iH^0t}\rt]}{\Tr_{{\cal H}}[\rho_0]},
\eeq
and we just have to take the trace by summing over the states (\ref{asA0}), and use the statement on every term (see Remark \ref{rematrace} for the subtleties in principle involved in taking the trace).

\begin{rema}
The result derived below can be seen as naturally related to the so-called Gell-Mann and Low theorem of QFT \cite{GML51}. This theorem essentially implies that if the limit on the left-hand side of (\ref{mapstates}) exist, then it must give the right-hand side. There are, however, many subtleties in the existence of this limit, and of the limit on the trace (\ref{toshow}) itself; in particular, the theorem itself involves an adiabatic process, where as here we have a quench. The arguments we present are based on the particular setup of the Keldysh formulation, with the particular order of limits involved, different from what is used in standard QFT constructions. Hence we believe it is worth going through the derivation without the explicit use of the Gell-mann and Low theorem.
\end{rema}

\begin{rema}\label{remaGGE}
In integrable systems, one expects that time evolutions from states that are not Hamiltonian eigenstates (quenches) give rise, after a large time, to Generalized Gibbs Ensembles \cite{GGE}. These ensembles are described by density matrices of the form $\rho_{\rm quench} = e^{-\sum_j \beta_j Q_j}$ where $Q_j$ are the local conserved charges (including the Hamitonian) and where the generalized (inverse) temperatures $\beta_j$ can be determined by requiring that the averages of the $Q_j$ before the start of the time evolution be reproduced by $\rho_{\rm quench}$. In the cases of integrable QFT, it is tempting to view our density matrix $\rho_{\rm stat}$ in this context. However, our result holds beyond integrability. Further, contrary to the usual quench situation, the density matrix $\rho_{\rm stat}$ does not reproduce the conserved averages of the charges $Q_j$. This is because $\rho_{\rm stat}$ reproduces averages of local fields and describes the steady-state region only, while conserved charges $Q_j$ are integration of local densities over the whole length of space including the far-away baths. The subtlety lies in the order of the limits that we must take: infinite length first, then infinite time.
\end{rema}

\subsubsection{Existence of steady state}

As a first step, we will show that the steady state exists, and that, in particular, we can replace the limit $t\to\infty$ by the Wick rotated limit,
\beq\label{ttau}
	\lim_{t=-i\tau,\; \tau\to\infty},
\eeq
for the operator $e^{-iHt}e^{iH^0t}$ that's on the right of $\Or$ in (\ref{thh}) (and $t=i\tau$ for the one on the left). In order to show this, we write
\beq
	e^{-iHt} e^{iH^0t} = \Pexp\lt[
	i \int_0^{-t}d\kappa\, \delta H(\kappa)
	\rt]
\eeq
where $\delta H = H-H^0 = dx\, h(0)$ and $\delta H(\kappa) = e^{iH^0\kappa}\,\delta H \, e^{-iH^0\kappa}$. Then, we have
\beqa
	\frc{\Tr\lt[\rho_0 \,e^{-iH^0t} e^{iHt}
	 \Or e^{-iHt}e^{iH^0t}\rt]}{\Tr\lt[\rho_0\,e^{-iH^0t}e^{iHt}e^{-iHt}e^{iH^0t}\rt]} &=& 
	 \frc{\Tr\lt[\rho_0\,\Pexp\lt[
	i \int_{-t}^0d\kappa\, \delta H(\kappa)
	\rt]\;\Or\;
	\Pexp\lt[
	i \int_0^{-t} d\kappa\, \delta H(\kappa)
	\rt]\rt]}{\Tr\lt[\rho_0\,\Pexp\lt[
	i \int_{-t}^0d\kappa\, \delta H(\kappa)
	\rt]
	\Pexp\lt[
	i \int_0^{-t} d\kappa\, \delta H(\kappa)
	\rt]\rt]} \n
	&=&
	\lt\bra\Pexp\lt[
	i \int_{-t}^0d\kappa\, \delta H(\kappa)
	\rt]\;\Or\;
	\Pexp\lt[
	i \int_0^{-t} d\kappa\, \delta H(\kappa)
	\rt]\rt\ket_0^{\rm connected}
	\label{pexp}
\eeqa
Then, we use the fact that for local fields in QFT, we have the form
\beq
	\bra \Or_1(0) \Or_2(\kappa) \ket_0^{\rm connected}
	\sim \kappa^\# e^{-\# T|\kappa|} e^{im\kappa} \quad\mbox{as}\quad |\kappa|\to\infty
\eeq
and similar equations for time separation of groups of fields. We use this with the local fields being $\Or$ and various products of $\delta H(\kappa)$ at various times $\kappa$. This implies that the integrals in (\ref{pexp}) all converge as $t\to\infty$, hence existence of the steady state. Further, in the path-ordered exponential on the right we can replace $t\mapsto -i\tau$ (and then in the integral $\kappa = ik$) and we get convergent integrals still (opposite shift on the left), with the same value; hence we can use (\ref{ttau}).

\subsubsection{Vacuum mapping}

Since the asymptotic states (\ref{asA0}) are constructed from the vacuum $|\vac\ket^0$, we start by describing the effect of $e^{-iHt} e^{iH^0t}$ on $|\vac\ket^0$. We employ a standard argument from elementary QFT, used in textbooks in order to obtain, for instance, the perturbative series from the operator formalism.

Let us write $|\vac\ket^0$ as a linear combination of eigenstates of $H$:
\[
	|\vac\ket^0 = B_\vac |\vac\ket + \sum_{{\rm excited\ states\ }s}
	B_s |s\ket =
	B_\vac |\vac\ket + \int d\theta \sum_\alpha B_\alpha(\theta) |\theta\ket_\alpha + \ldots
\]
Then
\[
	e^{-iHt} e^{iH^0t}|\vac\ket^0 = e^{-iHt}|\vac\ket^0
	= B_\vac|\vac\ket + \sum_{{\rm excited\ states\ }s}
	B_s\,e^{-iE_st}|s\ket.
\]
We have shown the large-$t$ limit can be taken by doing (\ref{ttau}). Then all contributions from the excited states vanish, and we get
\beq
	\lim_{t\to\infty} e^{-iHt} e^{iH^0t}|\vac\ket^0
	= B_\vac|\vac\ket.
\eeq
Hence, in the large-time limit, the vacuum $|\vac\ket^0$ collapses to $|\vac\ket$, as long as the overlap $B_\vac$ is nonzero (in which case it will normalize away by taking the ratio of traces in (\ref{toshow})).

In order to justify $B_\vac\neq0$, we use arguments similar to those of standard QFT textbooks, except for some changes due to the construction of the steady state requiring us to take the length of the system infinite first; the arguments then use in an essential way the locality of $\delta H = H-H^0$.

In standard QFT, in the context of perturbation theory, one compares a ``true'' vacuum $|\vac\ket$ with the vacuum of a free theory $|\vac\ket^0$. The differences between the Hamiltonian is a term integrated over the whole volume -- here the length of the system $\ell$ -- hence the overlap $\bra\vac|\vac\ket^0\sim e^{-\ell}$ decreases exponentially with $\ell$. Hence one usually assumes that the volume is finite, and take the limit of infinite volume afterwards; this gives rise to the usual perturbation theory. If we had a similar situation here, this would be very bad, since for the construction of the steady state, we need to take the limit $\ell\to\infty$ before $t\to\infty$. Happily, the difference between $H^0$ and $H$ is a local term (in fact, it is an infinitesimal term, but this is not the main point). This implies that the overlap $B_\vac = \bra\vac|\vac\ket^0$ is stable in the limit $\ell\to\infty$. Indeed, in the language of path integrals, we can represent it as
\[
	B_\vac = \frc{\bra\vac|\vac\ket^0}{\sqrt{\bra\vac|\vac\ket\;{}^0\bra\vac|\vac\ket^0}} =
	\frc{\int{\cal D}_{\low^0\cup \uH}\varphi e^{iS}}{
	\sqrt{\int{\cal D}_{\R^2} \varphi e^{iS}\;
	\int{\cal D}_{\R^0\times \R}\varphi e^{iS}}}
\]
where $\uH = \{(x,t):x\in\R,t\geq 0\}$. Let us divide space-time into cells of extent much larger than the inverse of the smallest mass, $m^{-1}$ (but still finite). Then the large-volume asymptotics of the path integral is obtained by factorizing it into such cells. Using time-reversal symmetry, we see that every cell, except three, cancel out in the ratio; in particular, for every cell, in the numerator, containing a space-time region ($t<0,\, x\approx 0$) cut into two pieces, there are two equal cells, in the denominator under the square root sign, containing a similar space-time region ($t<0,\, x\approx 0$ and $t>0,\,x\approx0$). Those not cancelled are those containing the space-time point $(0,0)$, because they are different in the three overlaps involved. Since only three cells remain, the result has a finite infinite-volume limit. Here we do not discuss UV divergencies: these are present, but one can UV regularize the theory and take the full scaling limit only at the end of the calculation of the nonequilibrium average (see Remarks \ref{remexlimits} and \ref{rematrace}).

\subsubsection{Excited states mapping}

Finally, in order to map the full excited states as in (\ref{mapstates}), we only need to show that (see (\ref{A}) and (\ref{A0}))
\beq\label{lla}
	\lim_{t\to\infty}
	e^{-iHt} e^{iH^0t} A_\alpha^0(\theta) e^{-iH^0t} e^{iHt}
	= A_\alpha(\theta).
\eeq
The main idea is to equate the ``asymptotic-state'' time $t=t_{asymp}$ used, in (\ref{A}) and (\ref{A0}), to construct the asymptotic states, with the negative $-t_{steady}$ of the ``steady-state'' time $t=t_{steady}$, used for instance in (\ref{lla}), to construct the steady state. This should be fine by the fact that the observable $\Or$ is of finite extent. Then (\ref{lla}) follows simply from
\beq\label{llab}
	e^{-iHt} e^{iH^0t} \h\phi^0_\alpha(x,-t) e^{-iH^0t} e^{iHt}
	=\h\phi_\alpha(x,-t).
\eeq

In order to justify equating $t_{asymp} = -t_{steady}$, consider the trace (\ref{thh}). We may use, instead of the asymptotic states, a set of states created as in (\ref{asA0}), but with $A_\alpha^0$ replaced by an expression like (\ref{A0}) where the $t$ limit has not been taken yet (hence neither the $L$ limit), fixing $t=t_{asymp}$. This set is not a basis of states, but it is likely that such states, at least for $t_{asymp}$ negative enough, span (a dense subset of) ${\cal H}^0$. Hence the trace can be written as a sum over such almost-asymptotic states, with appropriate normalization coefficients. Denoting $|s\ket_{t_{asymp}}^0$ the almost-asymptotic state corresponding to the asymptotic state $|s\ket^0$, and $C_s^0(t_{asymp})$ the associated coefficients (satisfying $\lim_{t_{asymp}\to-\infty} C_s^0(t_{asymp}) = 1$), we have formally
\beq
	\Tr_{{\cal H}^0}\big[\rho_0
	\,\cdot\big] \approx \sum_s w_s^0\,C_s^0(t_{asymp}) \;
	{}_{t_{asymp}}^{\hspace{0.72cm} 0}\bra s|\cdot|s\ket_{t_{asymp}}^0
\eeq
where $w_s^0$ is the weight (\ref{elr}). Denoting $|s\ket_{t_{asymp}}$ the similar almost-asymptotic states corresponding to (\ref{A}), and $C_s(t_{asymp})$, we have similarly
\beq
	\Tr_{{\cal H}}\big[\rho_{\rm stat}
	\,\cdot\big] \approx \sum_s w_s\,C_s(t_{asymp}) \;
	{}_{t_{asymp}}\bra s|\cdot|s\ket_{t_{asymp}}
\eeq
where $w_s$ is the weight (\ref{epm}). Now choosing $t_{steady} = -t_{asymp}$, Relation (\ref{llab}) and the equality (\ref{lrpm}) for the weights imply that we obtain, for the trace (\ref{thh}),
\beq
	\sum_s w_s\,C_s^0(-t_{steady}) \;
	{}_{-t_{steady}}\bra s|\Or|s\ket_{-t_{steady}}.
\eeq
For general $\Or$, not necessarily of finite extent, we couldn't expect the limit $t_{steady}\to\infty$ of this expression to give the trace on the right-hand side of (\ref{toshow}), because it is the different coefficient $C_s(t_{steady})$ that would be required. However, for observables of finite extent, we expect
\beq
	\lim_{t_{asymp}\to-\infty} {}_{t_{asymp}}\bra s|\Or|s\ket_{t_{asymp}} = \bra s|\Or|s\ket.
\eeq
Since the coefficient $C_s^0(-t_{steady})$ itself tends to 1, the large steady-state time can be taken, and we obtain (\ref{toshow}).

\begin{rema}
Note that it is crucial to take the limit of infinite system length first, as otherwise we clearly cannot replace the asymptotic-state time by the (negative of the) steady-state time (with finite lengths, the large time limit won't give asymptotic states).
\end{rema}

\subsection{A consequence on the average current}

The statement (\ref{toshow}) with (\ref{epm}) does not provide an immediate way of evaluating the average current or its fluctuations in general (in CFT it does, see the exact results in \cite{BD12}). Yet, an intuitive argument leads to the nice observation of left- and right- ``factorizaton'' of the average current. Clearly, the average energy current is
\[
	J = \frc{\Tr_{{\cal H}}\lt[\rho_{\rm stat} \,p(0) \rt]}{\Tr_{{\cal H}}[\rho_{\rm stat}]}
\]
where $p(x)$ is the momentum density. In cases where there is a sense in the notion of quasi-particles and quasi-momenta even in finite volume (for instance, in integrable models via the Bethe ansatz, see for instance \cite{FBA}), then there is a natural finite-length $L$ extrapolation $\rho_{\rm stat}^L$ of $\rho_{\rm stat}$. There is of course a finite-$L$ notion of $p$, and by translation invariance, we have $J = \lim_{L\to\infty} J_L$ with
\[
	J_L = L^{-1} \frc{\Tr_{{\cal H}_L}\lt[\rho_{\rm stat}^L P \rt]}{\Tr_{{\cal H}_L}[\rho_{\rm stat}^L]}
\]
where $P=\int_{-L/2}^{L/2} dx\, p(x)$ is the total momentum. Clearly, for every state $|v\ket$ with $n$ quasi-particles of momenta $p_j,\;j=1,\ldots,n$, we have
\[
	P|v\ket = \lt(\sum_{j=1}^n p_j \rt)|v\ket.
\]
This means that we can write $P=P_++P_-$ where $P_\pm$ are operators that measure the momenta of positive- and negative-momentum particles:
\[
	P_\pm|v\ket = \lt(\sum_{j:\;p_j{>\atop <} 0} p_j \rt)|v\ket.
\]
Since $\rho_{\rm stat}$ (and similarly $\rho_{\rm stat}^L$) factorizes into factors $\rho_{\rm stat,+}$ and $\rho_{\rm stat,-}$ acting on left- and right-moving particles, $\rho_{\rm stat} = \rho_{\rm stat,+} \rho_{\rm stat,-}$, similar to each other but depending on $\beta_l$ and $\beta_r$ respectively, we obtain
\beq\label{Jff}
	J = f(\beta_l) - f(\beta_r).
\eeq
The above derivation holds even if the quasi-particle / quasi-momentum description is not exact at $L$ finite, as long as it becomes ``more and more exact'' as $L\to\infty$, which should be the case for any near-critical or critical system.

The ``factorization'' relation (\ref{Jff}), as well as an equivalent factorization for the moment generating function, was established in the context of CFT in \cite{BD12}; further, interestingly, (\ref{Jff}) was observed numerically in \cite{KIM12} in spin chain models, hence beyond QFT\footnote{To my knowledge, Relation (\ref{Jff}) was first presented in my talk about the work \cite{BD12} with Denis Bernard at the Workshop on Non-equilibrium Physics and Asymmetric Exclusion Processes, ICMS, Edinburgh, 8 December 2011. There it was proposed in the context of general 1+1-dimensional QFT. It was also presented in my talk at the Galileo Galilei Institute workshop ``New quantum states of matter in and out of equilibrium'' on 23 April 2012, and appeared in the first version {\tt arXiv:1202.0239v1} of the preprint of \cite{BD12}.}. As was remarked in \cite{KIM12}, (\ref{Jff}) immediately implies a simple relation between the linear conductance and the nonequilibrium current: $J = J(T_l,T_r) = \int_{T_r}^{T_l} dT\,G(T)$ where $G(T) = d J(T',T)/dT'|_{T'=T}$.

\section{Concluding remarks}





We have shown how the Keldysh formulation leads, by QFT arguments, to a Gibbs-like ensemble formulation for nonequilibrium steady states of energy flow in quantum systems near to critical points.

We note that all the characteristics of the constructions were essential in order to obtain the proof: that the system's length be infinite, otherwise we cannot identify asymptotic-state and steady-state times, and that the quench be local (the term $\delta H = H-H^0$ is local), or at least of finite extent, otherwise $B_\vac$ vanishes. The existence proof of the steady state presented also uses both the infinite length and the locality (or finite extend) of $\delta H$.

Trying to generalize this construction to higher dimensions lead to interesting problems. If we have one additional dimension of finite length, then everything goes through essentially unchanged (except for subtleties in defining asymptotic states). However, as can be expected physically, an additional dimension of infinite length will cause problems. There, one would expect, immediately after connection of the two halves, to have an infinite total flow of energy, while in the steady state itself, to have a finite total flow of energy hence a vanishing energy current density. This slightly ``anomalous'' situation is translated into the vanishing of the overlap $B_\vac$: in the cell-cancellation argument presented, there remains infinitely many cells, leading to a vanishing proportional to $e^{-\ell}$ where $\ell$ is the extent of the additional dimension.

As we mentioned, the idea of certain Gibbs-like ensembles of asymptotic states representing a nonequilibrium steady state, and its link with the Keldysh formulation, has been used many times in the past. Hershfield's density matrix \cite{Hershfield93} was developed for charge transport, and was studied and used afterwards in the context of quantum impurities \cite{DA06,H07,D07,D09,DKHH11,BD12b}. Charge transport through impurities has also been studied using the asymptotic states of integrable QFT \cite{FLS95,KSL00,KSL01,BS07}, and using Bethe-ansatz asymptotic states \cite{MA07}, without explicit reference to a density matrix (although it is in some sense there implicitly). Nonequilibrium Gibbs ensembles of asymptotic states in spin chains were discussed with mathematical precision in \cite{R00,AP03,AJPP06} for heat transport, and were also studied in \cite{DSH11} in free fermion and boson models. Energy transport, including exact results for the current and the so-called full counting statistics, was studied in the context of CFT in \cite{BD12}, where a nonequilibrium density matrix was obtained in terms of the Virasoro algebra. The present notes give the first general study of energy transport in the context of massive relativistic QFT, providing the density matrix for nonequilibrium heat flows (which was proposed in \cite{BD12}) and a direct connection with the Keldysh formulation.

Much still needs to be done, of course. The evaluation of the energy current and its fluctuations is foremost: can concepts of integrability, combined with the present Gibbs-like ensemble description, be fruitfully used (I have work in progress with my collaborators on this subject)? Could one develop an efficient perturbation theory, including all the subtleties of renormalization, paralleling that for finite-temperature QFT? Deeper questions about the thermodynamic of nonequilibrium steady states may also be accessible from the present formulation. I hope that these notes will open the way to further developments in these directions.

\vspace{1cm}

{\bf Acknowledgements}

I am very much indebted to my collaborator Denis Bernard for many (still ongoing) discussions about the subject of nonequilibrium steady states, which ultimately led to my understanding of the situation in massive QFT. I also thank the participants to the Galileo Galilei Institute workshop ``New quantum states of matter in and out of equilibrium'' for questions raised on the subject which eventually led to this work, and I acknowledge in particular discussions with J. Bhaseen and F. Essler. I am of course very grateful to the students and lecturers of this Les Houches summer school for their insightful questions and comments, and in particular to Natan Andrei for comments and encouragement to write these notes.


\begin{thebibliography}{999}

\bibitem{BD12}
D. Bernard and B. Doyon, Heat flow in non-equilibrium conformal field theory, J. Phys. A: Math. Theor. 45 (2012) 362001, preprint
 {\tt arXiv:1202.0239}.

\bibitem{Lind}
G. Lindblad, On the generators of quantum dynamical semigroups, Commun. Math. Phys. 48 (1976) 119-130.

\bibitem{CaldeiraL81}
A.O. Caldeira and A.J. Leggett, Influence of Dissipation on Quantum Tunneling in Macroscopic Systems, Phys. Rev. Lett. {\bf 46}, 211 (1981).

\bibitem{SD07}
K. Saito and A. Dhar, Fluctuation theorem in quantum heat conduction, Phys. Rev. Lett. 99 (2007) 180601, preprint {\tt arXiv:cond-mat/0703777}.

\bibitem{SD11}
K. Saito and A. Dhar, Generating Function Formula of Heat Transfer in Harmonic Networks, Phys. Rev. E 83, 041121 (2011), preprint {\tt arXiv:1012.0622}.

\bibitem{Hershfield93}
S. Hershfield, Reformulation of steady state nonequilibrium quantum statistical mechanics, Phys. Rev. Lett. {\bf 70}, 2134 (1993).

\bibitem{R00}
D. Ruelle, Natural nonequilibrium states in quantum statistical mechanics, J. Stat. Phys. 98, 57 (2000), preprint {\tt arXiv:math-ph/9906005}.

\bibitem{AJPP06} W.H. Aschbacher, V. Jaksic, Y. Pautrat and C.-A. Pillet, Topics in non-equilibrium statistical mechanics, in {\em Open Quantum Systems III}, Springer, pp. 1-66, Aug. 2006, Lecture Notes in Mathematics.

\bibitem{C71}
C. Caroli, R. Combescot, P. Nozieres and D. Saint-James, Direct calculation of the tunneling current, J. Phys. C 4, 916 (1971).

\bibitem{JWM94} A.-P. Jauho, N. S. Wingreen and Y. Meir., Time-dependent transport in interacting and noninteracting resonant-tunneling systems, Phys. Rev. B 50, 5528 (1994), preprint {\tt arXiv:cond-mat/9404027}.

\bibitem{AP03} W.H. Aschbacher and C.-A. Pillet, Non-equilibrium steady states of the XY chain, J. Stat. Phys. 112 (2003) 1153.

\bibitem{DA06} B. Doyon and N. Andrei, Universal aspects of non-equilibrium currents in a quantum dot, Phys. Rev. B 73 245326 (2006), preprint {\tt arXiv:cond-mat/0506235}.

\bibitem{H07} J.E. Han, Mapping of strongly correlated steady-state nonequilibrium system to an effective equilibrium, Phys. Rev. B 75, 125122 (2007), preprint {\tt arXiv:cond-mat/0604583}.

\bibitem{D09} B. Doyon, The density matrix for quantum impurities out of equilibrium, lecture notes for the Fifth Capri Spring School on Transport in Nanostructure, {\tt http://tfp1.physik.uni-freiburg.de/Capri09}.

\bibitem{DKHH11} P. Dutt, J. Koch, J. Han, K. Le Hur, Effective equilibrium theory of nonequilibrium quantum transport, Ann. Phys. 326 (2011) 2963-2999, preprint {\tt  arXiv:1101.1526}.

\bibitem{MCP11} V. Moldoveanu, H.D. Cornean, C.-A. Pillet, Non-equilibrium steady-states for interacting open systems: exact results, Phys. Rev. B 84, 075464 (2011), preprint {\tt arXiv:1104.5399}.

\bibitem{BD12b} D. Bernard and B.Doyon, Full Counting Statistics in the Resonant-Level Model, J. Math. Phys. 53, 122302 (2012), preprint {\tt arXiv:1105.1695}.

\bibitem{DSH11} A. Dhar, K. Saito and P. Hanggi, Nonequilibrium density matrix description of steady state quantum transport, Phys. Rev. E 85, 011126 (2012), preprint {\tt arXiv:1106.3207}.

\bibitem{BLR00} F. Bonetto, J.L. Lebowitz and L. Rey-Bellet, Fourier's Law: a Challenge for Theorists, in {\em Mathematical Physics 2000}, edited by A. Fokas et. al. (Imperial College Press, London, 2000), p. 128, preprint {\tt arXiv:math-ph/0002052}.

\bibitem{LLP03} S. Lepri, R. Livi and A. Politi, Thermal conduction in classical low-dimensional lattices, Phys. Rep. 377, 1 (2003), {\tt arXiv:cond-mat/0112193}.

\bibitem{D08} A. Dhar, Heat Transport in low-dimensional systems, Adv. Phys. 57, 457 (2008), preprint {\tt arXiv:0808.3256}.

\bibitem{BD12c} D. Bernard and B. Doyon, Nonequilibrium steady states in conformal field theory, in preparation.

\bibitem{Sach} S. Sachdev, {\em Quantum Phase Transitions, Second Edition}, Cambridge University Press, New York, 2011.

\bibitem{D07} B. Doyon, New method for studying steady states in quantum impurity problems: the interacting resonant level model, Phys. Rev. Lett. 99 076806 (2007), preprint {\tt cond-mat/0703249}.

\bibitem{GML51} M. Gell-Mann and F. Low, Bound states in quantum field theory, Phys. Rev. 84, 350 (1951).

\bibitem{GGE} M. Rigol, V. Dunjko, V. Yurovsky and M. Olshanii, Relaxation in a completely integrable many-body quantum system: an ab initio study of the dynamics of the highly excited states of 1D lattice hard-core bosons, Phys. Rev. Lett. 98, 050405 (2007), preprint {\tt arXiv:cond-mat/0604476}.

\bibitem{FBA}  L.D. Faddeev, How algebraic Bethe ansatz works for integrable models, in Les Houches 1995, Relativistic gravitation and gravitational radiation pp. 149-219, preprint {\tt arXiv:hep-th/9605187}, 1996.

\bibitem{KIM12} C. Karrasch, R. Ilan, J.E. Moore, Nonequilibrium thermal transport and its relation to linear response, preprint {\tt arXiv:1211.2236}.

\bibitem{FLS95} P. Fendley, A.W.W. Ludwig and H. Saleur, Exact non-equilibrium transport through point contacts in quantum wires and fractional quantum Hall devices, Phys. Rev. B52 (1995) 8934, preprint {\tt arXiv:cond-mat/9503172}.

\bibitem{KSL00}  R. Konik, H. Saleur and A. Ludwig, Transport through Quantum Dots: Analytic Results from Integrability, Phys. Rev. Lett. 87, 236801 (2001), preprint {\tt arXiv:cond-mat/0010270}.

\bibitem{KSL01}  R. Konik, H. Saleur and A. Ludwig, Transport in Quantum Dots from the Integrability of the Anderson Model, Phys. Rev. B 66, 125304 (2002), preprint {\tt arXiv:cond-mat/0103044}.

\bibitem{BS07} E. Boulat and H. Saleur, Exact low temperature results for transport properties of the interacting resonant level model, Phys. Rev. B 77, 033409 (2008), preprint {\tt arXiv:cond-mat/0703545}.

\bibitem{MA07} P. Mehta and N. Andrei, Nonequilibrium Transport in Quantum Impurity Models: The Bethe Ansatz for Open Systems, Phys. Rev. Lett. 96, 216802 (2006), preprint {\tt arXiv:cond-mat/0508026}; P. Mehta, S.-P. Chao and N. Andrei, erratum, cond-mat/0703426.


\end{thebibliography}
\end{document}